# Poly(ionic liquid)-derived N-doped carbons with hierarchical porosity for lithium and sodium ion batteries

Walid Alkarmo,[a] Farid Ouhib,[a] Abdelhafid Aqil,[a] Jean-Michel Thomassin,[a] Jiayin Yuan,[b*] Jiang Gong,[c] Bénédicte Vertruyen,[d] Christophe Detrembleur,[a] Christine Jérôme [a,*]

[a] Centre for Education and Research on Macromolecules, CESAM Research Unit, University of Liege, Sart-Tilman B6a, 13allée du 6 août, B-4000 Liège, Belgium.

[b] Department of Materials and Environmental Chemistry, Stockholm University, Svante Arrheniusvag 16C, 10691, Stockholm Sweden.

[c] Key Laboratory for Material Chemistry of Energy Conversion and Storage, Ministry of Education, School of Chemistry and Chemical Engineering, Huazhong University of Science and Technology, Wuhan 430074, China.

[d] GREENMAT, CESAM Research Unit, University of Liège, Sart Tilman B6a, 4000 Liège, Belgium

*Corresponding authors.

E-mail address: jiayin.yuan@mmk.su.se, c.jerome@uliege.be

**Abstract:** The performance of lithium and sodium ion batteries relies notably on the accessibility to carbon electrodes of controllable porous structure and chemical composition. This work reports a facile synthesis of well-defined porous N-doped carbons (NPCs) using a poly(ionic liquid) (PIL) as precursor, and graphene oxide (GO)-stabilized poly(methyl methacrylate) (PMMA) nanoparticles as sacrificial template. The GO-stabilized PMMA nanoparticles were first prepared and then decorated by a thin PIL coating before carbonization. The resulting NPCs reached a satisfactory specific surface area of up to 561 $m^2/g$ and a hierarchically meso- and macroporous structure while keeping a nitrogen content of 2.6 wt %. Such NPCs delivered a high reversible charge/discharge capacity of 1013 mA h/g over 200 cycles at 0.4 A/g for lithium ion batteries (LIBs), and showed a good capacity of 204 mA h/g over 100 cycles at 0.1 A/g for sodium ion batteries (SIBs).

## 1. Introduction

In the past decades, nitrogen-doped porous carbons (NPCs) have attracted wide interest due to their potential in catalysis,[1] membrane separation,[2] $CO_2$ capture,[3] and energy technologies such as batteries,[4] fuel cells[5] and supercapacitors[6]. NPCs are particularly interesting for lithium ion batteries (LIBs) and very recently also for sodium ion batteries (SIBs).[7] The incorporation of N atoms favors some physical properties of NPCs by raising the overall electron density to enhance the electrochemical stability and charge mobility.[8] Furthermore, N dopant can carry surface functionality easily and a large number of defects to assist $Li^+$ or $Na^+$ insertion.[9,10] Meanwhile, pores in such a carbon matrix provide highways to $Li^+$ (or $Na^+$) diffusion and offer a large electrolyte/electrode interface for charge/mass transfer, thus enhancing specific capacity and rate performance.[11] Last but not least, these pores act as an ion reservoir and provide space to buffer volume expansion during $Li^+$ or $Na^+$ insertion/extraction to improve cycling stability.[12]

In general, NPCs are produced by pyrolysis of N-containing compounds such as synthetic polymers,[13] organic salts,[14] fossil fuels,[15] and biomass.[16] In terms of synthetic polymers, such as phenolic resins,[17] their easy access and processing are the major factors to be considered. In addition, some monomers such as pyrrole[18] may require metal catalysts for polymerization, leading to concerns of metal contamination. Moreover, extents of doping and graphitization, which affect electric conductivity, are important and can affect which polymer precursors are chosen. For example, the inherent nature of "hard carbon" precursors such as furfuryl and alcohol sucrose gives low graphitization degree at temperatures below 1200 ºC.[19] Numerous researchers in this field have confirmed that a broad range of properties of carbon materials depends on the chemical nature and morphology of the polymer precursors,[20–22] among which poly(ionic liquid)s (PILs) are an innovative class of carbon precursors.[23] In comparison to other polymers such as polyacrylonitrile, polypyrrole and polydopamine, PILs possess several unique features. Firstly, PILs present relatively high thermostability which leads to a high carbonization yield.[24] Secondly, they have diverse molecular structures carrying different heteroatoms such as nitrogen, boron, sulfur or phosphorus, which broadens the heteroatom-doping scope.[25] Thirdly, PILs are surface-active materials and are able to coat different surfaces (carbon, metal or metal oxide).[26] Consequently, PIL-derived porous carbons can be flexibly designed in different morphologies such as spheres, nanotubes and membranes, by templating or template-free methods.[27]

Recently, we prepared NPC composites with a nanostructured porosity *via* pyrolysis of polypyrrole deposited on the surface of GO nanosheet-stabilized PMMA particles that relied on the stabilizing power of GO nanosheets.[28-30] The resulting NPCs exhibited a specific surface area of 289~398 m²/g and a reversible capacity of 831 mAh/g at a current rate of 74.4 mA/g in LIBs. However, this reversible capacity decayed dramatically to 343 mAh/g at a higher rate of 744 mA/g.[28] In this contribution, poly(3-cyanomethyl-1-vinylimidazolium bromide) PIL as a popular carbon precursor,[13] was tested as NPC precursor in place of polypyrrole, which led to a NPC electrode of much better performance in LIBs and was further investigated as an SIB anode.

## 2. EXPERIMENTAL SECTION

The experimental details have been placed in the Supporting Information.

## 3. Results and discussion

As presented in Scheme 1, NPCs were synthesized by a templating method inspired by our previous work.[31] Briefly, a PMMA/GO hybrid template was prepared *via* dispersion polymerization of methyl methacrylate (MMA) in a water/methanol mixture in the presence of dispersed GO as stabilizer and 2,2'-azoisobutyronitrile (AIBN) as radical initiator.[30] During the polymerization, a phase separation occurred to form spherical particles of PMMA/GO with a particle size of 200~250 nm (Figure S2). The PMMA/GO template was then decorated by a thin layer of PIL as precursor for N-doped carbon. 3-Cyanomethyl-1-vinylimidazolium bromide (CMVImBr) was chosen as the IL monomer, as it is rich in nitrogen and is well soluble in the reaction mixture. By free radical copolymerization of this IL monomer and a divinylbenzene (DVB) crosslinker ([CMVImBr]/[DVB]=4/1 in molar ratio), a PIL thin film formed around the PMMA/GO particles in a suspension state, driven by a layer-by-layer effect. Scanning electron microscopy (SEM) analysis of the dried sample (Figure 1a) shows well-defined spherical particles of 250~300 nm in diameter. The strong interaction between the selected PIL and GO through non-covalent cation-π interactions as well as electrostatic attraction favors a homogeneous thin coating of PIL onto the surface of the preformed PMMA/GO particles.[32] The composition of these particles was investigated by Fourier transform infrared (ATR-FTIR) spectroscopy. Figure 1b compares the FTIR spectra of PMMA/GO and PMMA/GO/PCMVImBr. The characteristic peaks of PMMA/GO (C−O at 1147 cm$^{-1}$, C−O−C at 1435 cm$^{-1}$ and C=O at 1727 cm$^{-1}$, black line)[33,34] are all observed in that of PMMA/GO/ PCMVImBr (red line). Compared to the PMMA/GO spectrum, new bands at 1170 cm$^{-1}$ and 1553 cm$^{-1}$ in the PMMA/GO/PCMVImBr sample are observed and associated to the C–N stretching of imidazolium rings and ring in-plane asymmetric stretching, $CH_2(N)$ as well as $CH_3(N)CN$ stretching vibrations of the PIL side chains, respectively.[35,36] The FTIR analysis confirms that PIL was successfully grown onto PMMA/GO particles in accordance with the SEM images.

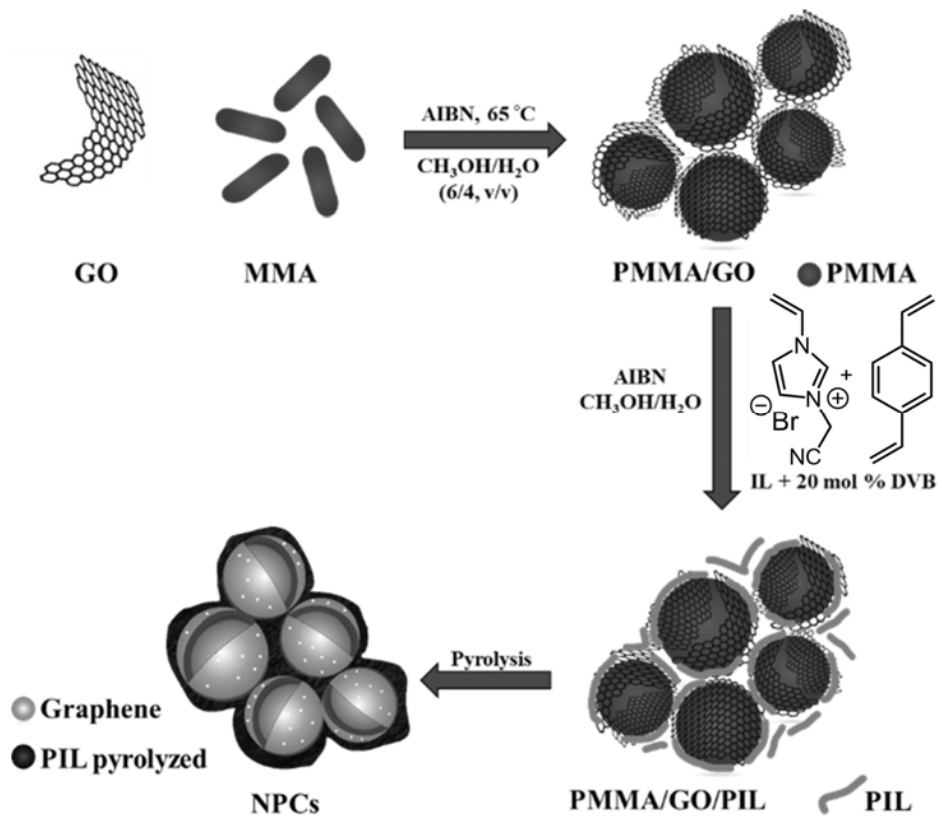

**Scheme 1.** Schematic illustration of the synthetic route towards NPCs from PIL, GO and PMMA nanoparticles.

The thermostability of the as-obtained PMMA/GO/PCMVImBr hybrid, *i.e.*, the NPC precursor, was checked by thermogravimetric analysis (TGA) under $N_2$ atmosphere and compared to the PMMA/GO template (Figure S3). The major mass loss of PMMA/GO sample (black line) occurs between 200 and 400 °C, which is attributed to a decomposition of oxygen containing groups in the GO and thermo-depolymerization of PMMA chains. A residue of 2.9 wt % was observed, which corresponds to carbon from the PMMA/GO sample. The PIL-containing NPC precursor (red line) has a much higher residue mass of 12.6 wt % at the same temperature.

Pyrolysis of PMMA/GO/PCMVImBr to generate NPCs was performed on a dry sample by a single-phase, stepwise thermal treatment. The first step was performed at 250 °C for 1 h under nitrogen to thermally depolymerize the sacrificial PMMA nanoparticle template, leaving behind macropores, and in parallel a thermal reduction of GO into graphene (rGO) nanosheets.[37] The second step involves pyrolysis of the PCMVImBr PIL network by heating at a rate of 5 °C/min until the desired temperature and then maintaining this temperature for 1 h. Two different final temperatures were used, *i.e.*, 800 and 900 °C,[38] yielding samples of NPC1 and NPC2, respectively. The morphology of the obtained NPCs was investigated by SEM (Figures 1c and d). Particle-like macroporous carbon products are clearly observed in both samples.

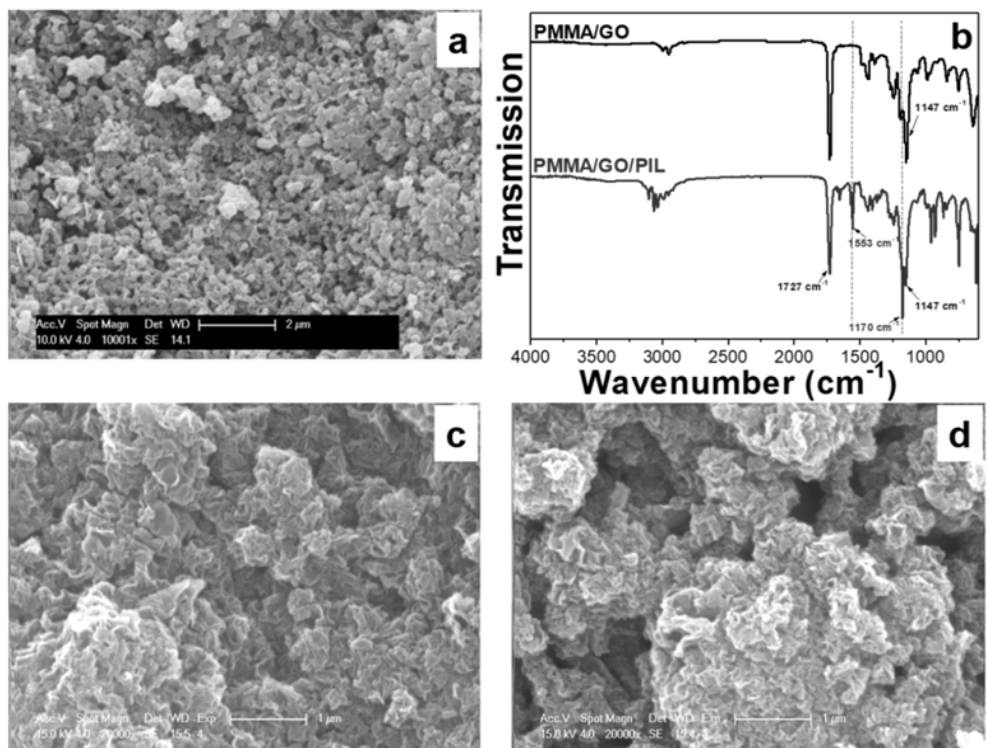

**Figure 1.** SEM images of (a) PMMA/GO/PIL, (c) NPC1, and (d) NPC2. (b) ATR-FTIR spectra of PMMA/GO, and PMMA/GO/PIL.

The crystalline phase of NPCs was studied by X-ray diffraction (XRD) measurement. Figure S4 shows the XRD patterns of the two (NPC1, and NPC2) samples. NPC1 (blue curve) exhibits two main diffraction peaks at $2\theta = 25.7°$ and $43.1°$,[39,40] which can be assigned, respectively, to the (002) and (100) planes of graphitic carbon. At a higher pyrolysis temperature to produce NPC2 (red curve), these two peaks appear at higher $2\theta$ angles, indicating the formation of graphitic carbons of a higher order in atomic structure.[41] The mass percentage of nitrogen in NPCs samples was determined by elemental analysis (Table 1) to be 2.9 and 2.6 wt %, respectively, which is normal for PILs carrying $Br^-$ as counter anion.[13,38]

The porous structure of NPCs was characterized by nitrogen sorption measurement at 77 K (Figures 2a and b). The Brunauer-Emmett-Teller specific surface area ($S_{BET}$) and the total pore volume are reported in Table 1. The hysteresis loops in the relative pressure of $P/P_0 > 0.45$ in the $N_2$ isotherms (Figure 2a) confirm the presence of mesopores, while the non-closure nature of the isotherm curve before $P/P_0$ -1.0 implies the co-existence of macropores that are generated during the degradation of PMMA nanoparticle template.[42,43] In addition, the $S_{BET}$ increases when NPCs are pyrolyzed at a higher temperature, *i.e.*, 424 m$^2$/g at 800 °C and 561 m$^2$/g at 900 °C. The increase in $S_{BET}$ might result from a higher extent of pyrolysis.[44] This presumption is supported by XRD measurement which indicates a better ordered graphitic structure at 900 °C than that at 800 °C. The pore size distribution curve of NPC1 and NPC2 samples was obtained by the non-local density functional theory (NLDFT) method (Figure 2b). A broad size distribution of mesopores are observed for both samples. One distinctive difference between these two samples is that the small fraction of micropores observed in NPC1 is completely absent in NPC2, *i.e.*, a full degradation of micropores at a higher pyrolysis temperature at the expense of growth of meso- and macropores.[44] This observation indicates that the micropores developed during the pyrolysis of PCMVImBr at 800 °C collapse and fuse into larger pores at 900 °C. The meso- and macropores observed for NPC1 and NPC2 originate from PMMA nanoparticle template degradation and the PIL precursor. It should be noted that the pyrolysis of PCMVImBr alone failed to a produce porous carbon structure.[13] Because a wide size distribution of meso-/macropores in a hierarchical manner is expected to be beneficial to metal ion batteries, the NPC samples were integrated in a half-cell for electrochemical tests.

**Table 1.** Characteristics of NPC1 and NPC2 obtained at 800 and 900 °C, respectively.

| Sample | Temperature of pyrolysis (°C) | N (wt %) | $S_{BET}$ (m²/g) | $V\mu$ (cm³/g) |
|---|---|---|---|---|
| NPC1 | 800 | 2.9 | 424 | 1.07 |
| NPC2 | 900 | 2.6 | 561 | 1.60 |

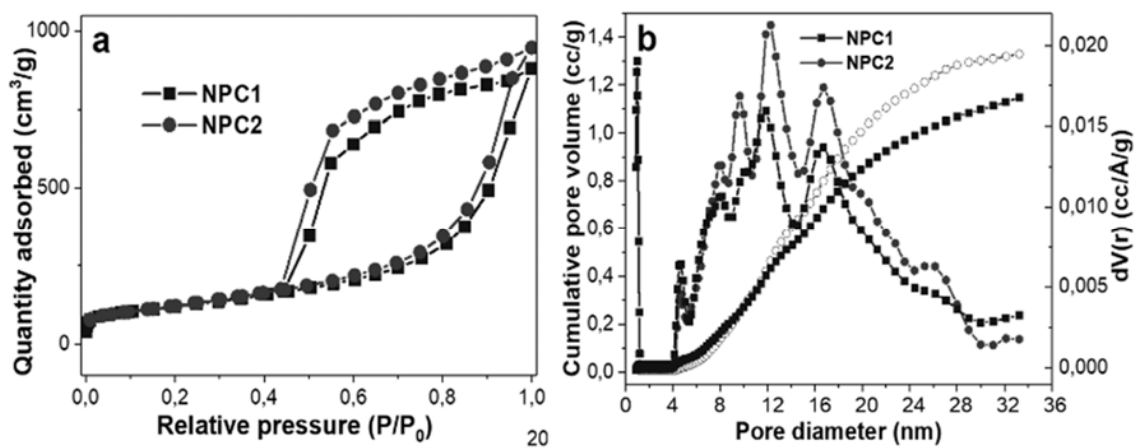

**Figure 2.** (a) Nitrogen adsorption-desorption isotherms and (b) pore size distribution plots of NPC1 and NPC2.

Several electrochemical properties of NPCs were first investigated by cyclic voltammetry (CV) in a half-cell configuration at a scanning rate of 0.1 mV s$^{-1}$ between 0 and 3.0 V (*vs.* Li$^+$/Li). As shown in Figures 3a and b, both NPC electrodes present typical CV curves of porous carbon materials.[40,45] The large shoulder peak between 1 and 0.4 V in the first cathodic scan, which disappears in the other cathodic scans, can be attributed to side reactions on the electrode surface and interface due to formation of solid-electrolyte interphase (SEI) layer.[46] No apparent peak was recorded in anodic scan, indicating that lithium ion extraction from NPC has no specific voltage. It is important to note that the CV curves after the first scan almost overlapped with each other, indicating a stable reversibility in both NPC1 and NPC2 electrodes.

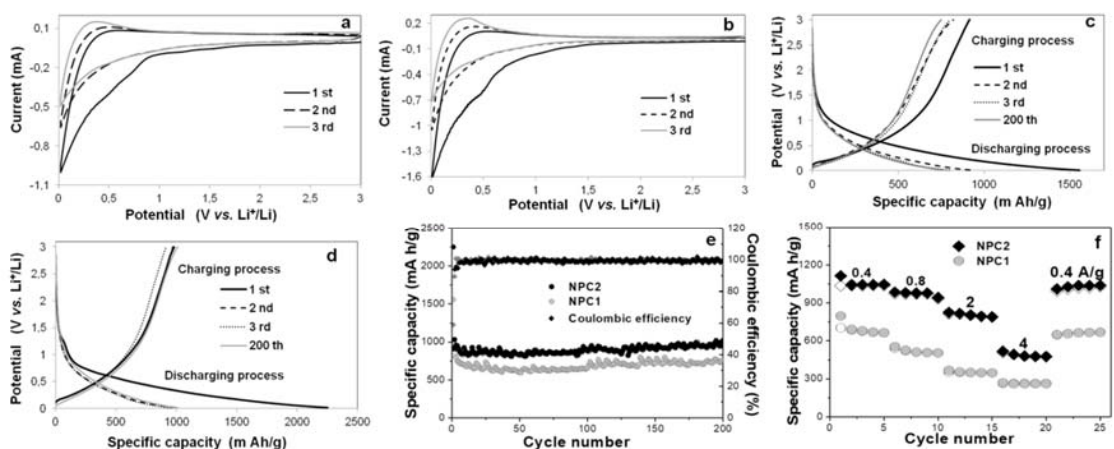

**Figure 3.** Cyclic voltammetry curves at a scan rate of 0.1 mV s$^{-1}$ of (a) NPC1, and (b) NPC2 electrodes. Charge-discharge curves measured at 0.4 A/g of (c) NPC1, and (d) NPC2 electrodes. (e) Cycle performance of the cells at a current rate of 0.4 A/g between 3.0 and 0 V vs Li$^+$/Li: specific capacity (circles) and Coulombic efficiency (diamonds) of the NPC1 (pink) and NPC2 (green) electrodes. (f) Rate capabilities and cycle performance of the NPC1 (circles) and NPC2 (diamonds) electrodes at current densities from 0.4 to 4 A/g: discharge capacities (colored marks) and charge capacities (hollow marks).

Figures 3c and d present galvanostatic charge-discharge profiles of the first three cycles and the 200$^{th}$ cycle for the NPC1 and NPC2 electrodes/Li half-cell at a current density of 0.4 A/g between 0 and 3.0 V. The first cycle reveals a high discharge capacity of 1557, and 2251 mA h/g and a charge capacity of 916 and 981 mA h/g for NPC1 and NPC2, respectively, representing initial coulombic efficiencies of 58.8% for NPC1 and 43.6% for NPC2. This irreversible capacity can be mainly ascribed to consumption of Li$^+$ in the formation of the SEI layer at the electrode-electrolyte interface and/or irreversible lithium ion insertion into highly active sites such as defects and in the vicinity of residual hydrogen atoms.[45,47,48] However, the irreversible capacity is reduced in the second cycle and the reversible capacity reaches 764 mA h/g for NPC1 and 1013 mAh/g for NPC2 after 200 cycles (Figure 3e). The coulombic efficiency of both samples increases dramatically upon cycling, and reaches 99.1% for NPC1 and 98.2% for NPC2 after 5 cycles, indicating an improved reversibility of the Li$^+$ ion intercalation and deintercalation stability. Figure 3f presents the rate capability performances of NPC1 and NPC2 electrodes at various current densities from 0.4 to 4 A/g. As depicted in Figure 3f, the rate performance of the NPC2 is superior to NPC1. The NPC2 electrode delivers reversible charge/discharge capacities of 1046, 942, 790, 478 and 423 mA h/g, at current rates of 0.4, 0.8, 2 and 4 A/g, respectively. Then the reversible charge/discharge capacity is recovered to 1039 mA h/g when the current density is set back to 0.4 A/g, indicating an excellent rate capability of NPC2. In comparison to NPC1 or to the reported N-doped carbon derived from polypyrrole prepared in the same manner, NPC2 exhibits excellent reversible capacity, high rate performances, and prolonged cycling stability.[28] The improved performance can be related to a higher specific surface area (561 m$^2$/g) and hierarchically meso- and macroporous structure, offering a sufficient electrode-electrolyte interface to exchange Li ions.[49,50] Moreover, the inherent N dopant (2.6 and 2.2 wt % as shown in Table 1) in the carbon matrix is beneficial for enhanced reactivity and creates more Li$^+$ storage sites.[61] Table S1 reports a comparison between the lithium storage performances of NPC2 and other N-doped carbons reported previously. It clearly shows that NCP2 delivers comparable or higher electrochemical performance than the best reported N-doped carbons in the literature.[53,55,62] This work shows that the nature of the polymer precursor for NPC and the pyrolysis temperature affect the structure characteristics of NPCs, such as specific surface area, the structured porosity and N-doping level, and are effective ways to improve cycling performance of electrode for LIBs.[18,40,51–53]

The NPC2 electrode was further investigated as an anode for sodium storage in a half-cell conformation with sodium metal counter electrode and NaTFSI as electrolyte. Figure 4a displays charge/discharge profiles of NPC2 electrode in 1st, 2nd, 3rd and 100th cycles between 0 and 3.0 V (vs $Na^+$/Na) at 0.1 A/g. The specific discharge capacity in the first cycle is 566 mA h/g, while the specific charge capacity is 329 mA h/g, defining a Coulombic efficiency of 58%. This large irreversible capacity is mainly attributed to formation of a SEI layer, similar to the LIBs. The Na cell cycling performance in a potential window of 0-3.0 V at a current rate of 0.1 A/g is presented in Figure 4b. After the initial cycle, the Coulombic efficiency increases to 99.8%. The reversible discharge capacities of NPC2 electrode after 100 cycles are maintained at 205 mA h/g, demonstrating good electrochemical performance of NPC2 electrode for SIBs. Similar to the LIBs, the high specific surface area and the hierarchically porous structure of NPC2 make an active surface accessible to the electrolyte and shortens the diffusion path for Na ions.[65–68] The capacities of SIBs are lower than LIBs due to the much bigger atomic radius of sodium ion ($Na^+$ 1.02 Å vs $Li^+$ 0.59 Å),[69] which leads to smaller discharge capacities due to poorer kinetics in SIBs compared to those in LIBs. The electrochemical performance of our NPC2 electrode is nevertheless satisfactory among candidates of anodes in SIBs reported recently.[70,71]

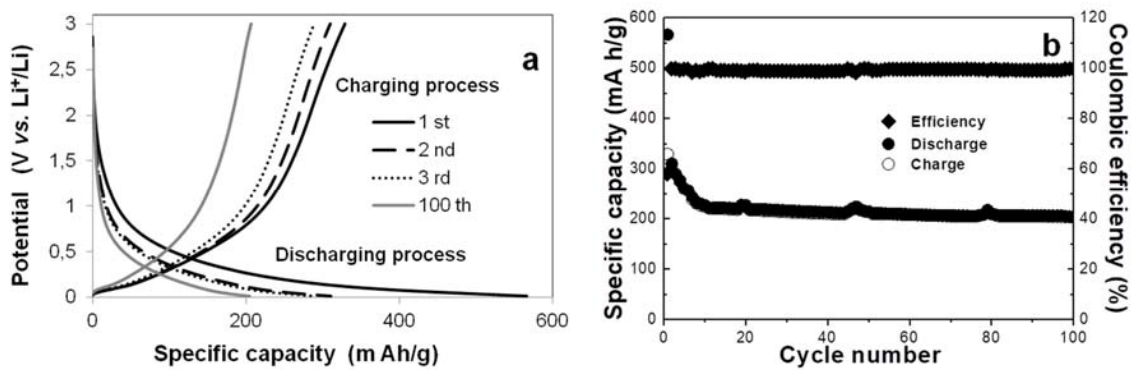

**Figure 4.** (a) Charge-discharge curves and (b) Cycling performance: specific capacity (circles) and coulombic efficiency (diamonds) of NPC2 electrode at a current density of 0.1 A/g between 0-3 V vs $Na^+$/Na.

## 4. Conclusions

In summary, NPCs were successfully prepared using PIL as carbon precursor and PMMA nanoparticles stabilized by GO as sacrificial template. PIL provides the nitrogen source of NPCs and facilitates the formation of hierarchical pores. The correlation between pyrolysis temperature and the nature of NPCs was preliminarily investigated in terms of specific surface area and porosity. NPC2 obtained at 900 °C shows a larger specific surface area of 561 m$^2$/g with hierarchical meso- and macropore structure and a nitrogen content of 2.6 wt %. More importantly, the NPC2 electrode displays a high reversible charge/discharge capacity of 1013 mA h/g after 200 cycles at 0.4 A/g for LIBs, and 205 mA h/g after 100 cycles at 0.1 A/g for SIBs. This good electrochemical performance clearly demonstrate that such NPCs are promising anode materials for both LIBs and SIBs. It is believed that without any complex techniques or post-treatments, this inexpensive strategy opens up a new stimulating platform for developing novel nitrogen-doped porous carbons from PIL for applications in energy storage, catalysis and environmental treatment.

**Acknowledgements**

The authors thank the ITN Marie-Curie "Renaissance" funded by the People FP7 Programme, the "Fonds de la Recherche Scientifique" (FRS-FNRS) and the Belgian Science Policy in the frame of the Interuniversity Attraction Poles Program (P7/05)-Functional Supramolecular Systems (FS2) for financial supports. C.D. is Research Director of the FRS-FNRS. J. Yuan thanks the Wallenberg Academy Fellow program of the Knut & Alice Wallenberg Foundation.

# Supporting information

## 1. Experimental details

### 1.1. Materials.

Methyl methacrylate (MMA; > 99%, Aldrich) and α,α′- azoisobutyronitrile (AIBN) (Fluka) were used without further purification. Graphene oxide solution (GO; N002-PS: 0.5 wt % in water, thickness 1.0-1.2 nm, x−y dimension ∼100 nm) was purchased from Angstron Materials. 1-Vinylimidazole (99%), bromoacetonitrile (97%) and divinylbenzene (DVB) were obtained from Sigma-Aldrich. 3-Cyanomethyl-1-vinylimidazolium bromide (CMVIZBr) was prepared according to literature procedures.[1,2] The methanol and other materials (Aldrich) were used as received.

### 1.2. Synthesis of PMMA/GO template

PMMA nanoparticles were prepared *via* the free radical polymerization of methyl methacrylate in presence of GO as stabilizer.[3] In a typical run, MMA (1 mL, 9.388 mmol) and AIBN (30 mg, 0.182 mmol) were added in a flask containing a mixture of 10 mL of methanol and 6 mL of an aqueous solution of GO (0.5 wt %). The mixture was degassed with flowing $N_2$ for 2 min and then stirred vigorously at 1000 rpm for 1h at 60 °C. After 30 minutes, the coloration turned from black to brown with the appearance of PMMA/GO particles. These particles are then used as a sacrificial template for preparing the nitrogen-doped porous carbon (NPC).

### 1.3. Synthesis of PMMA/GO/PIL precursor

The suspension of PMMA/GO in the methanol/water (6/4, v/v) mixture was sonicated by an ultrasonic device for 5 min. Then, CMVIZBr monomer (150 mg, 0.701 mmol) dissolved in deionized water (1 mL), 18.2 mg of DVB as crosslinker (20 mol %) and 3 mg of AIBN were added to the PMMA/GO suspension under vigorous stirring (1000 rpm). The dispersion was degassed by purging with nitrogen for 15 min prior to polymerization. The reaction was conducted under vigorous stirring at 60 °C overnight. The PMMA/GO/PIL particles were separated from the dispersion by centrifugation (10000 rpm for 10 min) and washed with deionized water. After three successive centrifugation/redispersion cycles with deionized water, the PMMA/GO/PIL particles were then dried at 80 °C for 12 h under vacuum until constant weight.

### 1.4. Pyrolysis process

In a typical experiment, the dried PMMA/GO/PIL samples were placed into an aluminum oxide crucible in an oven and heated under a $N_2$ atmosphere to 800 or 900 °C following the sequence as schematized in Figure S1: (i) during 1 h nitrogen was insufflated in the oven chamber at room temperature, (ii) then heated during 1 h to 250 °C and (iii) the temperature was kept at 250 °C during 1 h for degrading the sacrificial template, (iv) then heated during 1 h to 800 or 900 °C, and finally (v) the temperature was kept at 800 or 900 °C during 1 h for pyrolysis of PIL. Then the sample was cooled down slowly to room temperature. The obtained samples were denoted as NPC.

### 1.5. Characterization

The morphology of the samples was examined by scanning electron microscopy (SEM; JEOL JSM 840-A) with an accelerating voltage of 15 kV under high vacuum after metallization with Pt (30 nm). In addition, Transmission electron microscopy (TEM) was carried out on a Zeiss EM 912 Omega microscope operating at 120 kV. For sample preparation, one drop of the sample dispersion was placed on a 200 mesh carbon coated copper grid and dried in air. Thermogravimetric analysis (TGA) was carried out using a Q500 machine from TA instruments under a $N_2$ atmosphere at a heating rate of 5 °C/min. The structure of the precursor was studied by Fourier transform infrared spectroscopy (FT-IR, PerkinElmer Spectrum BX FTIR instrument) spectra. X-ray diffractograms were collected in Bragg-Brentano geometry using Cu K$_{alpha}$ radiation (Bruker D8 Twin-Twin). The N-doping levels of the samples was checked by elemental analysis on a Vario Micro setup. The N content reported is the average of four measurements. Nitrogen adsorption-desorption measurements were performed to evaluate the porosity of the materials. The specific surface area, $S_{BET}$, was calculated using the Brunauer-Emmett-Teller (BET) equation, with the adsorption data taken in the relative pressure range 0.01 to 0.10. Pore size distribution was determined by the non-local density functional theory (NLDFT) method.

### 1.6. Electrochemical measurements

Electrochemical characterization of the NPC materials as anode in rechargeable lithium and sodium ion batteries was performed at room temperature in a half-Li or Na coin cells (CR2025 type) by galvanostatic. The working electrodes were fabricated by grinding NPCs as active materials, carbon black as conducting agent and polyvinylidene fluoride (PVDF) dissolved in N-Methyl-2-pyrrolidone (NMP) as binder in a mass ratio of 8:1:1 to form a homogeneous slurry. Then, the working electrodes were prepared by slurry casting on a copper foil, and then dried in a vacuum oven at 80 °C for 12 h to remove completely the excess of solvent. Pure metal lithium (LIBs) or sodium (SIBs) foils were used as both reference and counter electrodes and a microporous polypropylene membrane (CelgardP) was used as separator. The electrolyte used in the coin cells was composed of 90 μl of 1M LiPF6 in a mixture of LP71 (1 M LiPF6 in EC:DEC:DMC 1:1:1) for LIBs and 1M NaTFSI in EC:DMC (1:1 vol.%) 99.9% for SIBs (Merck). Cycling tests were carried out with voltages between 0.01 V and 3.0 V (vs. Li$^+$/Li or vs. Na$^+$/Na) using a Biologic VMP3 multichannel potentiostat. The specific capacity of the anodes was calculated based on the total mass of the active materials.

## 2. Supporting Figures

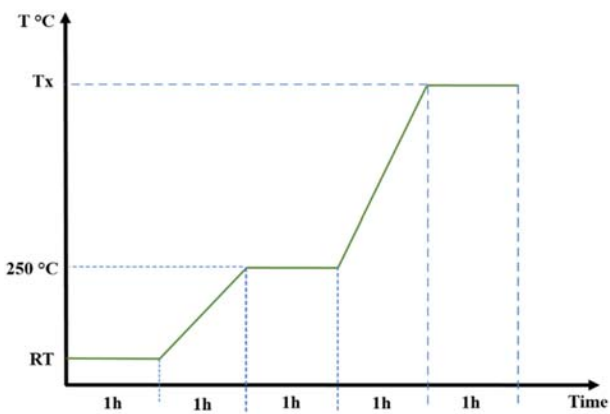

**Figure S1.** Pyrolysis temperature program of PMMA/GO/PIL precursor. $T_x$: 800 or 900 °C.

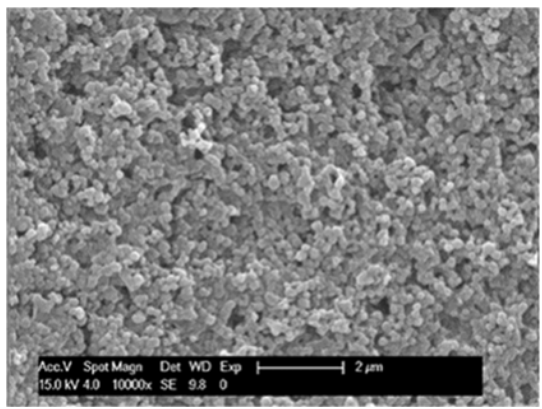

**Figure S2.** SEM micrograph of PMMA/3 wt% GO particles.

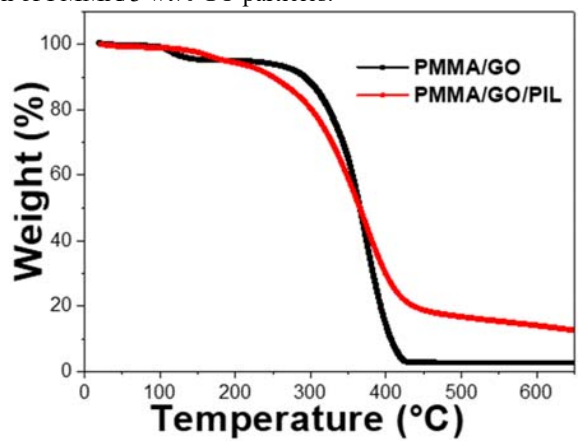

**Figure S3.** TGA thermograms of PMMA/GO and PMMA/GO/PIL.

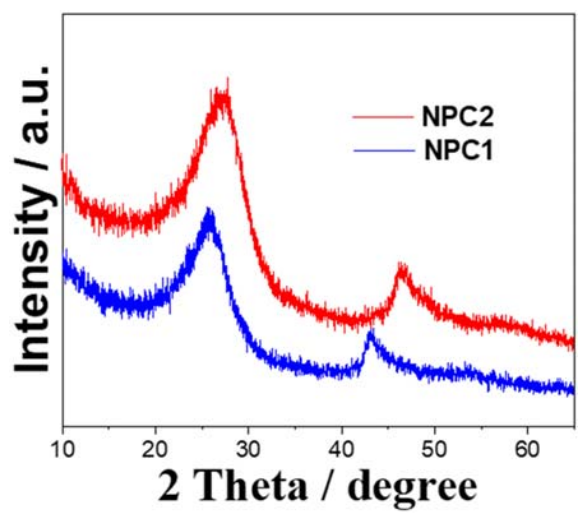

**Figure S4.** XRD patterns of NPC1 and NPC2 samples.

**Table S1.** Comparison of the electrochemical performance of various N-doped carbon-based anodes materials.

| N-doped carbon anodes | Current density (mA/g) | Reversible capacity (mA h/g) | Voltage window (V vs. Li/Li$^+$) | Initial coulombic efficiency (%) | Nitrogen content | Specific surface area (m$^2$/g) | Ref. |
|---|---|---|---|---|---|---|---|
| N-porous carbon nanofibers | 50 | 1323 | 0.01-3.0 | 58.9 | 7.9 at% | 1198 | 53 |
| N-doped carbon fibers | 30 | 576 | 0.005-3.0 | 82.8 | 12.6 wt% | 381 | 54 |
| N-doped graphene | 50 | 1177 | 0.01-3.0 | - | 2.1 at% | - | 55 |
| Mesoporous N-doped carbon | 100 | 1780 | 0-3.0 | 55 | 10.1 wt% | 805.7 | 56 |
| N-doped porous graphene | 100 | 672 | 0.005-3.0 | 62 | 5.8 at% | 1170 | 57 |
| N-doped graphene nanoribbons | 100 | 714 | 0.01-3.0 | 63 | 3.7 at% | - | 58 |
| N-doped graphene sheets | 100 | 832.4 | 0.05-3.0 | 44.8 | 19.5 at% | 504 | 59 |
| N-doped carbon sponge | 500 | 870 | 0.001-3.0 | 34.7 | 2.38 at% | 613 | 60 |
| N-doped carbon spheres | 500 | 540 | 0.01-3.0 | 60 | 5.43 wt% | 67.4 | 18 |
| N-doped graphene frameworks | 200 | 700 | 0.005-3.0 | 52.3 | 2.6 wt% | 610 | 61 |
| N-doped porous carbon materials | 100 | 488 | 0.005-3.0 | 62.8 | 9.75 at% | 482.06 | 7 |
| N-carbon/rGO nanosheets | 100 | 1100 | 0.003-3.0 | 58.3 | 15.4 wt% | 327 | 62 |
| N-enriched carbon nanofibers | 50 | 400 | 0.01-3.0 | 85 | 24.4 at% | - | 63 |
| N-doped graphene | 50 | 550 | 0.01-3.0 | 61.2 | 3.15 at% | 127.1 | 64 |
| **N-doped porous carbon network (NPC2)** | **400** | **1013.0** | **0.01-3.0** | **43.6** | **2.58 wt%** | **561** | **Current work** |